\documentstyle{article}\begin{document}
\title{Relativistic diffusive motion in thermal electromagnetic fields  }
\author{Z. Haba \\
Institute of Theoretical Physics, University of Wroclaw,\\ 50-204
Wroclaw, Plac Maxa Borna 9, Poland\\email:zhab@ift.uni.wroc.pl}
\date{}\maketitle
\begin{abstract}
We discuss  relativistic dynamics in a random electromagnetic
field which can be considered as a high temperature limit of the
quantum electromagnetic field in a heat bath (cavity) moving with
a uniform velocity  $w$. We derive a diffusion approximation for
particle's dynamics generalizing the diffusion of Schay and
Dudley. It is shown that J\"uttner distribution is the equilibrium
state of the diffusion.
\end{abstract}

 \section{Introduction} The dynamics of a test (tracer)
particle in a medium of some other particles is often approximated
by diffusion \cite{pitaj}.  It is an old result
\cite{to}\cite{plasma} that the Vlasov equation
 describing a tracer particle  moving
 in a random electromagnetic field produced by a chaotic
 motion of other particles
  can be approximated by a diffusion equation.
 The diffusion approximation finds
applications  in heavy ion collisions
\cite{svetitsky}\cite{rafelski}, plasma physics and in
astrophysics \cite{to}\cite{plasma}\cite{jop}\cite{stu}. A
detailed discussion of the diffusion approximation  can be found
in \cite{kubo1}\cite{kubo2} \cite{kampen}\cite{kesten} \cite{fri}.

The mathematical theory of a relativistic diffusion is not so well
developed as the non-relativistic one (see the reviews in
\cite{hang}\cite{deb}). A relativistic diffusion in the
configuration space does not exist \cite{lopuch}\cite{hakim}. If
the diffusing particle has a fixed mass then an analog of the
Kramers diffusion (on the phase space) without friction is
uniquely determined. The generator of the diffusion is defined by
the Laplace-Beltrami operator on the mass shell as has been shown
by Schay \cite{schay} and Dudley \cite{dudley}. In our earlier
papers \cite{habapre}\cite{habagr}we have studied a friction term
in the framework of the relativistic Brownian motion of
Schay,Dudley, Franchi and Le Jan \cite{lejan}. The friction term
has been defined as a drift uniquely determined by the equilibrium
state and by the requirement of the reversibility of the process
(in physical terms by the detailed balance). Such a drift term has
been  obtained by other means in \cite{dunkel}\cite{calo}. Without
a friction the relatvistic Brownian motion is unphysical,e.g., the
particle energy grows exponentially fast \cite{habagr}.

Following a development of the theory of the non-relativistic
Brownian motion we may expect that relativistic deterministic
motion, which is chaotic in certain sense, can also be approximated
by a relativistic diffusion. In our earlier paper \cite{habajpa} we
have  discussed the dynamics of a relativistic particle in a
Poincare invariant random electromagnetic field. We have shown that
the  relativistic diffusion of Schay and Dudley can be considered as
an approximation of the proper time dynamics in this field. However,
such a Poincare invariant electromagnetic field is rather
unphysical. In this paper we discuss particle's dynamics in a
thermal (random) electromagnetic field with a distinguished frame of
reference. We can think of the Universe filled with the CMB
radiation as an example of a Poincare covariant system with a heat
bath \cite{cmb}. When describing a random system it is more
appropriate to apply a statistical (Liouville) description. The
 evolution in the proper time  is determined by the  expectation value of an exponential
  of the Liouville operator. The expectation value is defined by the
  probability distribution of the electromagnetic field. The
  two-point function is expressed by its Fourier transform
  $\tilde{G}$.
 We do not need to specify the thermal distribution $\tilde{G}$ (it can be
the Planck distribution). We consider an observer moving in an
inertial frame with the velocity $w$ (this is equivalent to moving
the heat bath and keeping the observer at rest). For such an
observer the thermal distribution is a function of the momentum
$k$ of the electromagnetic modes and $w$
\cite{cmb}\cite{kujawski}. We restrict ourselves to the expansion
of the exponential of the Liouville operator till the second order
in the electromagnetic field strength (the square of the Liouville
operator). Assuming that the correlation functions of the
electromagnetic modes (Fourier components) are decaying fast for
the time larger than the correlation time $\tau_{c}$  we can apply
the Markov approximation treating the particle's subsequent
scatterings upon electromagnetic modes as independent. According
to Kubo's  theory \cite{kubo1}-\cite{kubo2} the second order
differential operator arising in the small time expansion is the
generator of the diffusion defining the evolution of the system
also at times large in comparison to the correlation time
$\tau_{c}$.  We show that the form of this diffusion is determined
uniquely by the assumptions of the Lorentz invariance of
$\tilde{G}$. This assumption  allows to express the expectation
value of the square of the Liouville operator unequivocally up to
two constants. It follows that the generator of the diffusion
depends only on the momenta and on the frame $w$. The diffusion
equation in the proper time is a generalization of the diffusion
of Schay \cite{schay} and Dudley \cite{dudley} (corresponding to
zero temperature). The particle mass is preserved by the dynamics.
It enters in an additive way in the model. We can sum over the
mass when computing expectation values of observables, e.g., the
energy-momentum or entropy. We think that keeping the mass as a
variable (we can fix it at the end if necessary)  can be a useful
tool in some models of diffusion. In quantum physics \cite{wert}
the Wigner function is an analog of the classical phase space
distribution . The square of four-momentum is off the mass shell
in the Wigner function description \cite{wert} (a matter with a
continuous mass spectrum has been suggested recently in
\cite{georgi}). In an application of the diffusion equation to
cosmology \cite{bert}\cite{cosmo} the mass of a galaxy can be
considered as a parameter varying continuously. The diffusion
equation, which we have obtained in a model of an interaction with
the electromagnetic thermal radiation, could have a broader
application to any diffusive dynamics in an environment
characterized by  a distinguished reference frame (the heat bath).
The diffusion in the proper time originated from the dynamics in a
random field. In a random system the  meaning of the proper time
is loosing its physical meaning. We should express the evolution
of states and observables (e.g., the energy-momentum) in the
coordinate (laboratory) time. We obtain a transport equation in
the laboratory time from the requirement that probability density
should not depend on the proper time.
 There is a certain similarity between our transport equation
and the one of refs.\cite{malik}\cite{chev}\cite{chev2}. These
authors describe a diffusion of a particle in a fluid. The fluid
supplies a distinguished reference frame. Our form of the equation
and the constants appearing in it result from a concrete physical
model which has not been considered in
refs.\cite{malik}-\cite{chev2}.

The plan of the paper is the following. In secs.2-3 we repeat the
formulation \cite{habajpa} of the kinetic equation in an
electromagnetic field in order to fix the notation. In sec.4 the
two-point function of a random electromagnetic field  depending
additionally  on the four-velocity $w^{\mu}$ is discussed. The
exponential of the Liouville operator is expanded  till the second
order in the proper time. Its expectation value is calculated.  In
sec.5 we take the diffusion generator as calculated in previous
sections as a basis of the relativistic diffusion theory. We show
that J\"uttner distribution is the equilibrium state of the
diffusion.

\section{Kinetic equation of particles in an electromagnetic field}

 The  dynamics of a
particle in an electromagnetic field is described by the equations
\cite{landau}
 \begin{equation}
 \sqrt{p^{2}}\frac{dx^{\mu}}{d\tau}=p^{\mu},
 \end{equation}
 \begin{equation}
\sqrt{p^{2}}\frac{dp_{\mu}}{d\tau}={\cal F}_{\mu\nu}p^{\nu},
 \end{equation}
where $\mu=0,1,2,3$, $x$ denotes the coordinate and $p$ the
four-momentum .  ${\cal F}$ is an antisymmetric field strength of
the electromagnetic field. We could express the mass as
$\sqrt{p^{2}}$ without changing the meaning of eqs.(1)-(2) because
it follows from eq.(2) that
$p^{2}=\eta_{\mu\nu}p^{\mu}p^{\nu}=const$ as a consequence of the
antisymmetry of ${\cal F}_{\mu\nu}$ and
\begin{equation} \eta^{\mu\nu}p_{\mu}\frac{d}{d\tau}p_{\nu} =0
\end{equation} with
$\eta_{\mu\nu}=(1,-1,-1,-1)$ . From eqs.(1)-(2) it follows that
$\tau$ is proportional to the proper time as
\begin{displaymath} d\tau^{2}=dx^{\mu}dx_{\mu}=
 c^{2}dt^{2}(1-c^{-2}(\frac{d{\bf x}}{dt})^{2}),
 \end{displaymath}
 where $x^{0}=ct$ ( we
set $c=1$ from now on for convenience).  We can eliminate $\tau$
from eqs.(1)-(2) in favor of $x^{0}$ if we wish.

In the kinetic theory of classical  particles  we consider a
distribution ${\cal K}_{\tau}$ of trajectories
$(x(\tau,y,q),p(\tau,y,q))$ in the  phase space
 starting from $(y,q)$
\begin{equation}
{\cal
K}_{\tau}(x,p;y,q)=\delta(x-x(\tau,y,q))\delta(p-p(\tau,y,q)).
\end{equation} We can consider more general
distributions of trajectories $\Phi_{\tau}$ by spreading the
initial points in the phase space with a certain probability
distribution $\Phi$
\begin{equation}
\Phi_{\tau}(x,p)=\int {\cal K}_{\tau}(x,p;y,q)\Phi(y,q)dydq.
\end{equation}
${\cal K}_{\tau}$ as well as $\Phi_{\tau}$ satisfy the
differential equation
\begin{equation}\begin{array}{l}\frac{d}{d\tau}\Phi_{\tau}\equiv {\cal G}^{+}\Phi_{\tau}=
-\frac{1}{\sqrt{p^{2}}}(p^{\mu}\frac{\partial\Phi_{\tau}}{\partial
x^{\mu}}+{\cal F}^{\mu\nu}p_{\nu}\frac{\partial\Phi_{\tau}}{\partial
p^{\mu}}).
\end{array}\end{equation} ${\cal K}_{\tau}$ with  the  initial condition
$\delta(x-y)\delta(p-q)$ and  $\Phi_{\tau}(x,p)$ with the initial
condition $\Phi(x,p)$ . We defined here a first order operator
${\cal G}^{+}$ which will be discussed later in more detail.  In
eq.(6) we assumed that $p_{0}$ is an independent variable. Working
with the four-momentum off the mass shell allows to preserve
explicit Lorentz invariance. A formulation with an independent
$p_{0}$ may have applications to quantum phase space methods based
on the Wigner function \cite{wert}\cite{elze}\cite{mrow}. If $p$ is
on the mass shell then we skip $p_{0}$ derivative in eq.(6).

The density of trajectories $\Omega$ in the laboratory time
$x^{0}$ can be expressed by $\Phi_{\tau}$ \cite{wert}
\begin{equation}
\Omega(x,p)=\int d\tau \Phi_{\tau}(x,p).
\end{equation}
It satisfies the transport equation \begin{equation} {\cal
G}^{+}\Omega=0.
\end{equation}In the Liouville approach to the classical statistical
mechanics we consider functions $W$ on the phase space as
observables. We may define the expectation value $\Phi_{\tau}(W)$
of $W$ in the state $\Phi_{\tau}$ as
\begin{equation}\begin{array}{l}\Phi_{\tau}(W)=
(\Phi_{\tau},W)=\int
dxdp\Phi_{\tau}(x,p)W(x,p)\equiv(\Phi,W_{\tau})\cr= \int
dydq\Phi(y,q)W(x(\tau,y,q),p(\tau,y,q)).\end{array}
\end{equation}   The semigroup $(T_{\tau}W)(y,q)\equiv
W_{\tau}(y,q)=W(x(\tau,y,q),p(\tau,y,q))$ is unitary in the scalar
product (9). We have from the definition of $W_{\tau}$  in eq.(9)
\begin{equation}\begin{array}{l}
\frac{d}{d\tau}(T_{\tau}W)(y,q)=(T_{\tau}{\cal G}W)(y,q) =({\cal
G}T_{\tau}W)(y,q) =\frac{1}{\sqrt{q^{2}}}(q^{\mu}\frac{\partial
W_{\tau}}{\partial y^{\mu}}+{\cal F}^{\mu\nu}q_{\nu}\frac{\partial
W_{\tau}}{\partial q^{\mu}})\cr={\cal G}W_{\tau},
\end{array}\end{equation}where the generator ${\cal G}$ of the
semigroup $T_{\tau}=\exp(\tau{\cal G})$ is the adjoint of ${\cal
G}^{+}$ of eq.(6) in the Hilbert space $L^{2}(dydq)$  (  by $dy$ or
$dq$ we denote an integral over $R^{4}$; vectors from $R^{3}$ will
be denoted by a boldface letter). In the derivation of eq.(10) we
apply the semigroup law $T_{\tau+s}=T_{s}T_{\tau}
 =T_{\tau}T_{s}$ in the form $W(x(s,x_{\tau},p_{\tau}),p(s,x_{\tau},p_{\tau}))
 =W(x(\tau,x_{s},p_{s}),p(\tau,x_{s},p_{s}))$ (where
 $(x_{\tau},p_{\tau})$ is the shorthand notation for the trajectory of
 eq.(4))
 resulting from the composition of trajectories. Then, the
 equality $T_{\tau}{\cal G}={\cal G}T_{\tau}$ in eq.(10) follows
 by differentiation. The requirement of the $\tau$-independence of
the probability distribution ($\frac{d}{d\tau}\Phi_{\tau}=0$) gives
the kinetic equation in the laboratory time $t$ . This is the same
equation as the one which can be derived by an elimination of $\tau$
in favor of $t$ in the evolution equations (1)-(2).

\section{Random dynamics}
In this  section we repeat some definitions used in our earlier
paper  \cite{habajpa} in order to fix the notation. We write the
formula (10) for an evolution of a function of trajectories in a
general form as
\begin{equation}
\frac{d}{d\tau} W=(X+Y)W,
\end{equation}
where
\begin{equation} X=\frac{1}{\sqrt{p^{2}}}p^{\mu}\partial^{x}_{\mu}
\end{equation}and \begin{equation}
Y=\frac{1}{\sqrt{p^{2}}}{\cal
F}^{\mu\nu}p_{\nu}\partial_{\mu}.\end{equation} From now on
derivatives over space-time coordinates will have an index $x$ and
the derivatives without an index mean the derivatives over
momenta. Let
\begin{equation}
Y(\tau)=\exp(-\tau X)Y\exp( \tau X) =\frac{1}{\sqrt{p^{2}}}{\cal
F}^{\mu\nu}(x-\frac{1}{\sqrt{p^{2}}}p\tau,p)p_{\nu}(\partial_{\mu}+\frac{1}{\sqrt{p^{2}}}\tau\partial_{\mu}^{x}).\end{equation}
Then, the solution of eq.(11) can be expressed in the form (an
analog of  the interaction picture in quantum mechanics)
\begin{equation}
W_{\tau}=\exp(\tau X)W^{I}_{\tau},
\end{equation}
where
\begin{equation}
\partial_{\tau}W^{I}_{\tau}=Y(\tau)W^{I}_{\tau}.
\end{equation}
We assume that the electromagnetic field is random. We consider an
expectation value of $W_{\tau}$ (we could equivalently discuss
$\Phi_{\tau}$ with the same result). In general, we have the
cumulant expansion for the expectation value (the expectation
value over the electromagnetic field ${\cal F}$ is denoted by
$\langle . \rangle$, the initial condition $W$ is assumed to be
independent of ${\cal F}$)
\begin{equation}\begin{array}{l}
\langle W_{\tau}^{I}\rangle=\exp\Big(\int_{0}^{t}ds \langle
Y(s)\rangle +\frac{1}{4}\int_{0}^{t}ds\int_{0}^{s}ds^{\prime}
\langle (A(s)A(s^{\prime})+A(s^{\prime})A(s))\rangle +....\Big)W,
\end{array}\end{equation}here
\begin{displaymath}
A(s)=Y(s)-\langle Y(s)\rangle \end{displaymath} If
$[Y(s),Y(s^{\prime})]=0$ and $Y$ is a linear function of Gaussian
variables then eq.(17) is exact (with no higher order terms).
 The expansion of the dynamics (11) till the second
order term reads
\begin{equation}\begin{array}{l}
\langle W_{\tau}^{I}\rangle=W+ \int_{0}^{\tau}ds\langle Y(s)\rangle
W+\frac{1}{2}(\int_{0}^{\tau}ds\langle Y(s)\rangle)^{2}W
+\frac{1}{2}\langle(\int_{0}^{\tau}ds A(s))^{2}\rangle
W+...\end{array}\end{equation}  The approach of Kubo
\cite{kubo1}-\cite{kubo2} approximates the random Liouville operator
in the exponential (17) by the $\tau^{2}$ term in the expansion
(18)(we discuss this approximation in the next section).
\section{Diffusion in random electromagnetic fields}
In ref.\cite{habajpa} we have calculated the generator of the
diffusion in a Poincare invariant random electromagnetic field. We
have assumed that the correlation functions of the electromagnetic
field depend only on the space-time coordinates. In this paper we
discuss the correlation functions which may depend also on a
time-like vector $w^{\mu}$ (without loss of generality we may
assume $w^{\mu}w_{\mu}=1$; as will be seen from the computations
the vector $w^{\mu}$ plays also a role of an ultraviolet cutoff).
We have in mind a model of a particle moving in the
electromagnetic field of the black body radiation at finite
temperature $\beta^{-1}$ (the heat bath). The Lorentz invariance
of the particle system is lost if we keep  the heat bath at rest.
We can express the Lorentz invariance of the whole system (a
particle interacting with the electromagnetic field and the
environment) considering a heat bath moving with a velocity
$w^{\mu}$.In quantum field theory the correlation functions of the
quantum electromagnetic field are usually calculated in the rest
frame of the heat bath \cite{kapusta}(for explicitly Lorentz
covariant calculations see \cite{weldon}). Correlation functions
of quantum fields are not symmetric (quantum fields do not
commute). Hence, they do not describe a random field. However, in
the high temperature limit the non-symmetric part can be neglected
\cite{kapusta}. In this limit the correlations can be represented
as correlation functions of random fields. We split

 \begin{equation}
 {\cal F}_{\mu\nu}= \langle{\cal F}_{\mu\nu}\rangle+F_{\mu\nu},
 \end{equation}
  where $F_{\mu\nu}$
 has zero mean value. We assume that an average $\langle\cdot \rangle$ over ${\cal F}$ is defined
which preserves the Poincare symmetry. This means that the
two-point function defined by
\begin{equation}\langle
F_{\mu\nu}(x)F_{\sigma\rho}(x^{\prime})\rangle=G_{\mu\nu
;\sigma\rho}(x-x^{\prime})
\end{equation}
is a tensor. $G_{\mu\nu ;\sigma\rho}$ is symmetric under the
exchange of indices $(\mu\nu;x)$ and $(\sigma\rho;x^{\prime})$ and
antisymmetric under the exchange $\mu\rightarrow\nu$ and
$\sigma\rightarrow\rho$. We impose the Bianchi identities ( as
${\cal F}$ is expressed by a potential)
\begin{equation}\partial_{\sigma}^{x}{ \cal F}_{\mu\nu,\sigma}+\partial_{\nu}^{x}{ \cal
F}_{\sigma\mu}+\partial_{\mu}^{x}{ \cal
F}_{\nu\sigma}=0.\end{equation} In terms of the two-point function
\begin{equation}
\partial^{x}_{\alpha}\epsilon^{\alpha\beta\mu\nu}G_{\mu\nu
;\sigma\rho}=0.
\end{equation}
For Fourier transforms eq.(20) reads
\begin{equation}\langle\overline{\tilde{F}}_{\mu\nu}(k)\tilde{F}_{\sigma\rho}(k^{\prime})\rangle=
\tilde{G}_{\mu\nu ;\sigma\rho}(k)\delta(k-k^{\prime}),
\end{equation}
where $\tilde{G}_{\mu\nu ;\sigma\rho}(k)$ is a tensor which must
be constructed from the vectors $k_{\mu}$,$w_{\mu}$ and the
fundamental four-dimensional tensors $\eta_{\mu\rho}$ and
$\epsilon_{\mu\nu\rho\sigma}$.  Hence,in general we could have
with some scalar functions $a_{r}$ (higher order tensors in $k$
and $w$ would not satisfy the antisymmetry requirements mentioned
below eq.(20))
\begin{equation}\begin{array}{l}
 \tilde{G}_{\mu\nu
 ;\sigma\rho}(k)=a_{1}(\eta_{\mu\sigma}k_{\nu}k_{\rho}-
 \eta_{\mu\rho}k_{\nu}k_{\sigma}+\eta_{\nu\rho}k_{\mu}k_{\sigma}-\eta_{\nu\sigma}k_{\mu}k_{\rho})
+a_{0}\epsilon_{\mu\nu\sigma\rho}\cr
+a_{2}(\eta_{\mu\sigma}w_{\nu}k_{\rho}-
 \eta_{\mu\rho}w_{\nu}k_{\sigma}+\eta_{\nu\rho}w_{\mu}k_{\sigma}-\eta_{\nu\sigma}w_{\mu}k_{\rho})\cr
+a_{3}(\eta_{\mu\sigma}w_{\nu}w_{\rho}-
 \eta_{\mu\rho}w_{\nu}w_{\sigma}+\eta_{\nu\rho}w_{\mu}w_{\sigma}-\eta_{\nu\sigma}w_{\mu}w_{\rho}).
\end{array}\end{equation}However, the Bianchi identities (22) lead to
$a_{2}=a_{3}=0$. The reason is that in eq.(22) (in the
$x$-representation) we need in addition to $\partial_{\alpha}^{x}$
one more derivative with an index $\nu$ or $\mu$ in order to give
zero on the rhs. There are terms multiplied by $a_{2}$ without such
a derivative,hence $a_{2}=0$. It is clear that the $a_{3}$ terms
cannot satisfy (22) because they are not symmetric in  $(\mu,\nu)$.
Hence, $a_{3}=0$. The requirement of positivity of the probability
measure defining the expectation value (20) leads to $a_{0}=0$ ( as
discussed in \cite{habajpa}; in the case of a quantum free
electromagnetic field at finite temperature \cite{kapusta} we obtain
also the representation (24) where only $a_{1}\neq 0$). Summarizing,
the two-point function of a random electromagnetic field satisfying
the Bianchi identities  must have the form
\begin{equation}\begin{array}{l}
 G_{\mu\nu
;\sigma\rho}(x,x^{\prime})=\int dk\tilde{G}_{\mu\nu
 ;\sigma\rho}(k)\exp(ik(x-x^{\prime}))\cr
 =\int dk \tilde{G}(k)\exp (ik(x-x^{\prime}))(\eta_{\mu\sigma}k_{\nu}k_{\rho}-
 \eta_{\mu\rho}k_{\nu}k_{\sigma}+\eta_{\nu\rho}k_{\mu}k_{\sigma}-\eta_{\nu\sigma}k_{\mu}k_{\rho}),
\end{array}\end{equation}
Hence,
\begin{equation} \langle
F_{\mu\nu}(x)F_{\sigma\rho}(x^{\prime})\rangle_{\beta}=-D_{\mu\nu
;\sigma\rho}G(x-x^{\prime}),\end{equation}
 where\begin{equation}\begin{array}{l}
 D_{\mu\nu;\sigma\rho}
=-\eta_{\mu\sigma}\partial^{x}_{\nu}\partial^{x}_{\rho}+
\eta_{\mu\rho}\partial^{x}_{\nu}\partial^{x}_{\sigma}-
\eta_{\nu\rho}\partial^{x}_{\sigma}\partial^{x}_{\mu}+
\eta_{\nu\sigma}\partial^{x}_{\mu}\partial^{x}_{\rho}.
\end{array}\end{equation}
   The two-point
function is positive definite if and only if $\tilde{G}(k)$ in
eq.(25) satisfies the condition

 \begin{equation} \tilde{G}(k)\geq 0
\end{equation} and $\tilde{G}(k)$=0 if $k^{2}< 0$ (as we have shown in \cite{habajpa}).

 It follows from eq.(25) that
\begin{equation}\begin{array}{l} (D_{\mu\nu;\sigma\rho}G)(x)
=\eta_{\mu\sigma}T_{\nu\rho}(x)- \eta_{\mu\rho}T_{\nu\sigma}(x)+
\eta_{\nu\rho}T_{\sigma\mu}(x)- \eta_{\nu\sigma}T_{\mu\rho}(x),
\end{array}\end{equation}
where \begin{equation}
T_{\mu\nu}(x)=-\partial^{x}_{\mu}\partial^{x}_{\nu}G(x)=\int
dk\tilde{G}k_{\mu}k_{\nu}\exp(ikx).
\end{equation}
We shall need solely \begin{equation}T_{\mu\nu}(0) =\int
dk\tilde{G}k_{\mu}k_{\nu}=\epsilon w_{\mu}w_{\nu}-\pi_{\epsilon}
(\eta_{\mu\nu}-w_{\mu}w_{\nu})\end{equation} The decomposition of
the rhs of eq.(31) follows from the Lorentz invariance. If
$\tilde{G}$ has the meaning of the distribution of electromagnetic
modes (photons) then $\epsilon$ is the energy density and
$\pi_{\epsilon}$ the pressure. The constants in eq.(31) can be
expressed by $\tilde{G}$
\begin{displaymath}
\epsilon=\int dk\tilde{G}(k)(kw)^{2},
\end{displaymath}\begin{displaymath}
\pi_{\epsilon}=\frac{1}{3}\int dk\tilde{G}(k)((kw)^{2}-k^{2}).
\end{displaymath}

 If $\tilde{G}(k) $ is vanishing for $k^{2}<0$ (as
is the case for a probability distribution of electromagnetic modes)
then
 from eq.(31) we can derive the inequality
 \begin{equation}
 \epsilon \geq 3\pi_{\epsilon}\geq 0.
 \end{equation}
We could express the expectation value $\langle
A(s)A(s^{\prime})\rangle$, needed for a calculation of the
diffusion, by
$\partial_{\mu}\partial_{\nu}G(\frac{1}{\sqrt{p^{2}}}(s-s^{\prime})p)$
of eq.(30)calculated at the point $(s-s^{\prime})p$
 ( it comes from the correlation functions of $
F^{\mu\nu}(x-\frac{1}{\sqrt{p^{2}}}ps,p)$ of eq.(14) calculated at
$s$ and $s^{\prime}$ with the covariance (26)). This function has
more involved tensor decomposition than $T^{\mu\nu}$ of eq.(31)
because there are still some tensors constructed from $p_{\mu}$
and $w_{\mu}$ on the rhs of eq.(30). However, these terms will be
of higher order than $\tau^{2}$. In the limit of small $\tau$ we
calculate
\begin{equation}\begin{array}{l}{\cal A}_{w}=\lim_{\tau\rightarrow
0}\frac{1}{\tau^{2}}\langle(\int_{0}^{\tau}A(s)ds)^{2}\rangle=
\lim_{\tau\rightarrow
0}\frac{1}{4\tau^{2}}\int_{0}^{\tau}ds\int_{0}^{s}ds^{\prime}\langle
A(s)A(s^{\prime})+A(s^{\prime})A(s)\rangle
\cr=(-\eta^{\mu\sigma}T^{\nu\rho}(0)+\eta^{\mu\rho}T^{\nu\sigma}(0)-\eta^{\nu\rho}T^{\mu\sigma}(0)
+\eta^{\nu\sigma}T^{\mu\rho}(0))
\frac{1}{\sqrt{p^{2}}}p_{\nu}\partial_{\mu}\frac{1}{\sqrt{p^{2}}}p_{\rho}\partial_{\sigma}.
\end{array}\end{equation}
We dropped the $s\partial^{x}$ term from eq.(14) because it does not
contribute to the limit of small $\tau$. If we apply the formula
(31) then we obtain
\begin{equation}\begin{array}{l}
{\cal A}_{w}\equiv 2\pi_{\epsilon}
P^{\mu\nu}\partial_{\mu}\partial_{\nu}-6p^{-2}\pi_{\epsilon}
p^{\mu}\partial_{\mu}\cr
-(\epsilon+\pi_{\epsilon})p^{-2}((pw)^{2}\partial^{\mu}\partial_{\mu}-(wp)(w^{\mu}p^{\rho}+w^{\rho}p^{\mu})
\partial_{\mu}\partial_{\rho}+p^{2}w^{\mu}w^{\rho}\partial_{\mu}\partial_{\rho})
\cr+(\epsilon+\pi_{\epsilon})p^{-2}(p^{\nu}\partial_{\nu}+2wpw^{\nu}\partial_{\nu})\cr
\equiv\partial_{\mu}\alpha^{\mu\nu}\partial_{\nu},\end{array}\end{equation}where

\begin{equation}
\alpha^{\mu\nu}=2\pi_{\epsilon} P^{\mu\nu}
-(\epsilon+\pi_{\epsilon})p^{-2}((pw)^{2}-(wp)(w^{\mu}p^{\nu}+w^{\nu}p^{\mu})
+p^{2}w^{\mu}w^{\nu})=P^{\mu\sigma}C_{\sigma\rho}P^{\rho\nu}
\end{equation}
with \begin{equation}
p^{2}P^{\mu\sigma}=\eta^{\mu\sigma}p^{2}-p^{\mu}p^{\sigma}
\end{equation}
and \begin{equation} C_{\sigma\rho}=2\pi_{\epsilon}
\eta_{\sigma\rho}- (\epsilon+\pi_{\epsilon})(
(wp)^{2}p^{-2}\eta_{\sigma\rho}+w_{\sigma}w_{\rho}).
\end{equation}   It should be pointed out that without the
velocity $w$ the limit in eq.(33) would not exist. In fact,
$T_{\mu\nu}(0)$ is infinite if $\tilde{G}$ depends only on
$k^{2}$. A damping factor in $\tilde{G}$ is necessary  if the
integral (31) is to exist (such a damping factor is present in the
Planck distribution (47)).

 The approach of Kubo
\cite{kubo1}-\cite{kubo2} approximates the random Liouville
operator on the rhs of eq.(18) by the $\tau^{2}$ term.  Kubo shows
\cite{kubo1}-\cite{kubo2} that the kinematic  $\tau^{2}$ behaviour
in random dynamics at times short in comparison to the correlation
time $\tau_{c}$ (of the electromagnetic modes) goes into the
diffusive $\tau$ behaviour at times large in comparison to the
correlation time. Kubo's argument can be considered as a Markov
approximation ( no memory of successive steps) to the random
dynamics.

Another interpretation of the diffusion approximation (which can be
made rigorous \cite{kesten}) interprets the small $\tau$ expansion
as a weak field expansion. The weak field can be rescaled to a large
time. The rescaling can be interpreted as a transformation of a
microscopic  behaviour of a particle system into an evolution at  a
macroscopic time scale. In more detail , eq.(2) is rewritten as
\begin{displaymath}
\frac{dp^{\mu}}{d\tau}=\delta^{2}\frac{1}{\sqrt{p^{2}}}F^{\mu\nu}p_{\nu}.
\end{displaymath}
Then, rescaling the time and an expansion in $\delta$ lead to the
diffusion with the diffusion tensor \begin{equation}
\tilde{\alpha}^{\mu\sigma}= \int_{0}^{\infty}ds\langle
F^{\mu\nu}(x)F^{\sigma\rho}(x-s\frac{1}{\sqrt{p^{2}}}p)\rangle
p_{\nu}p_{\rho}.
\end{equation}The integral (38) is proportional to the correlation
time $\tau_{c}$. As discussed in \cite{kubo1}-\cite{kubo2}  even if
the correlations of the random fields are not decaying exponentially
then nevertheless the behaviour of the particle dynamics in random
fields described by the matrix $\alpha$ for times small in
comparison to the correlation time determines the behaviour for
times large in comparison with the correlation time. The large time
behaviour is diffusive with the diffusion matrix $\tilde{\alpha}$
related to $\alpha$ in eq.(34) by
\begin{equation}
\tilde{\alpha}^{\mu\sigma}=\tau_{c}\alpha^{\mu\sigma}.
\end{equation}
The Kubo's formulas (the analogs of eqs.(38)-(39)) are widely
applied in the transport theory of random systems.

 As will be discussed in sec.5 in general the diffusion generated by
${\cal A}_{w}$  (34) has no equilibrium. We need a friction term
in order to stabilize the behaviour at large time. We would like
to point out that if $\langle \partial_{\sigma}^{x}{\cal
F}_{\mu\nu}\rangle \neq 0$ then in the expansion (18) there is a
drift term which can lead to a friction . Let us consider in
eq.(18) the term which is of the first order in the field.
Expanding ${\cal F}(x-s\frac{1}{\sqrt{p^{2}}}p)$ in $s$ we obtain
\begin{equation}\begin{array}{l}
K=\int_{0}^{\tau}ds\langle Y(s)\rangle
 =\tau\langle {\cal F}_{\mu\nu}(x)\rangle \frac{1}{\sqrt{p^{2}}} p^{\nu}\partial^{\mu}
 -\tau^{2}\langle \partial^{x}_{\sigma}{\cal F}_{\mu\nu}\rangle p^{-2}
p^{\sigma}p^{\nu}\partial^{\mu}+o(\tau^{2}).\end{array}
\end{equation}Poincare invariance implies that
\begin{equation}
\langle{\cal F}_{\mu\nu}\rangle=0 \end{equation} because no
antisymmetric tensor can be built from
$\epsilon^{\mu\nu\alpha\beta}$, $\eta^{\mu\nu}$ and $w^{\mu}$ .
Assuming that the second term on the rhs of eq.(40) is non-zero we
have from the Poincare invariance
\begin{equation} \langle
\partial_{\sigma}^{x}{\cal
F}_{\mu\nu}\rangle=-\frac{r}{2}(\eta_{\sigma\mu}w_{\nu}-\eta_{\sigma\nu}w_{\mu})
\end{equation}with a certain constant $r$. Hence,
\begin{equation}
 K=rwpp^{-2}p_{\mu}\partial^{\mu}-rw_{\mu}\partial^{\mu}=-rP^{\mu\nu}w_{\mu}\partial_{\nu}.
\end{equation}with
$P^{\mu\nu}=\eta^{\mu\nu}-p^{-2}p^{\mu}p^{\nu}$.
We can get an interpretation of eq.(42) if we calculate an
electric current of the electromagnetic field
\begin{equation} {\cal J}_{\nu}=\partial_{x}^{\mu}{\cal F}_{\mu\nu} .\end{equation}
From eq.(42)
\begin{equation}
\langle {\cal J}^{\mu}\rangle= -\frac{3r}{2} w^{\mu}.
\end{equation}
The current (45) results from a charge $-\frac{3r}{2}$
 moving with the velocity
$w^{\mu}$. We can obtain such a current  in a finite temperature
quantum field theory of interacting electromagnetic and complex
scalar fields $\phi$ \cite{kapusta}. The density matrix is
\begin{equation}
\rho_{\beta}=\exp(-\beta P^{\nu}w_{\nu}),
\end{equation} where $\beta^{-1}$ is the temperature,$P_{\mu}$ is the four-momentum of the quantum
fields and $w^{\mu}$ describes the moving frame. Then, calculating
the expectation value (of the electric current) we obtain in the
lowest order of the expansion in the coupling constant
\begin{equation}\begin{array}{l}\langle {\cal J}_{\mu}\rangle_{\beta}=- iTr\Big(\exp(-\beta
P^{\nu}w_{\nu})(\overline{\phi}\partial_{\mu}^{x}\phi-\phi\partial_{\mu}^{x}\overline{\phi})\Big)
 \cr=-\int d{\bf k}k_{0}^{-1}k_{\mu}(\exp(\beta kw)-1)^{-1}=-\frac{3r}{2} w_{\mu}
 \end{array}\end{equation} with a certain constant $r$. Here, $\langle . \rangle_{\beta}$
 denotes an expectation value (defined by the trace)
  with respect to the state (46) .

\section{Relativistic diffusion at finite temperature}
We define the diffusion generator by the lowest order (
$\tau^{2}$) term in the expansion (18) in proper time (including
the friction term coming from a constant current as discussed at
the end of sec.4). We incorporate Kubo's  argument that this
operator multiplied by the correlation time generates a diffusion
properly approximating the random dynamics. In this way we are led
to the following diffusion equation for particles with a
continuous mass spectrum ( by a certain abuse of notation we
denote by $W$ the expression which has the meaning of $\langle
W\rangle $ in random dynamics of sec.3)
\begin{equation}\begin{array}{l}
\frac{1}{\tau_{c}}\partial_{\tau}W\equiv {\cal
G}_{w}W=\frac{1}{\tau_{c}}p^{\mu}\partial^{x}_{\mu}W+\partial_{\mu}\alpha^{\mu\nu}\partial_{\nu}W
-rP^{\mu\nu} w_{\nu}\partial_{\mu}W,\end{array}\end{equation}
where $r $ is determined by the expectation value (42).

 The
diffusion in the proper time originates from the relativistic
dynamics in the proper time. Then, the proper time has a
well-defined meaning. This is the time measured in the frame moving
with the particle. However, if the particle motion is random then
this frame is random as well. Such a random frame is loosing a
physical meaning. For this reason  we should express the kinetic
transport in terms of the coordinate (deterministic,laboratory)
time. The probability distribution evolves according to the adjoint
equation (6). The independence of the proper time parametrization
($\frac{d}{d\tau}\Phi_{\tau}=0$) is equivalent to the replacement of
the proper time by the physical time $t$ in the kinetic equation
(8). We extend this requirement to the diffusion (48). In this way
we obtain the transport equation
\begin{equation}
{\cal
G}_{w}^{+}\Omega=-\frac{1}{\tau_{c}}p^{\mu}\partial_{\mu}\Omega
+{\cal A}_{w}\Omega+r\partial_{\mu}P^{\mu\nu}w_{\nu}\Omega=0,
\end{equation} where the adjoint of ${\cal G}_{w}$ is  in $L^{2}(dxdp)$ .
 Explicitly,
\begin{equation}\begin{array}{l}
\frac{1}{\tau_{c}}p^{\mu}\partial^{x}_{\mu}\Omega=\partial_{\mu}(\alpha^{\mu\nu}\partial_{\nu}+r
P^{\mu\nu}w_{\nu})\Omega\equiv\partial_{\mu}\Big(2\pi_{\epsilon}
P^{\mu\rho}\partial_{\rho} +rP^{\mu\rho} w_{\rho}\cr
-(\epsilon+\pi_{\epsilon})p^{-2}((pw)^{2}\partial^{\mu}-(wp)(w^{\mu}p^{\rho}
+w^{\rho}p^{\mu})\partial_{\rho}+p^{2}w^{\mu}w^{\rho}\partial_{\rho}\Big)\Omega.
\end{array}\end{equation}
 If
the condition (32) is satisfied then
$\partial_{\mu}\alpha^{\mu\nu}\partial_{\nu}$ is a non-positive
operator as an expectation value (33) of the square of a
differential operator. It follows  (see e.g.,\cite{operators})
that the momentum dependent coefficients $\alpha^{\mu\nu}$ satisfy
the positivity condition
\begin{equation}a_{\mu}a_{\nu}\alpha^{\mu\nu}\geq 0.\end{equation}
However, we can prove that the inequality (51)  is satisfied under
the  condition
\begin{equation}\epsilon\geq \pi_{\epsilon}\geq 0\end{equation}
weaker than (32). We can use the Lorentz invariance of the
inequality (51) in order to choose a proper Lorentz frame. So, if
$a^{2}>0$ we choose a frame such that $a=(a_{0},{\bf 0})$. In this
frame (from eq.(35)) we have \begin{equation}
a_{\mu}a_{\nu}\alpha^{\mu\nu}=a_{0}^{2}p^{-2}
((\epsilon-\pi_{\epsilon}){\bf p}^{2}+(\epsilon+\pi_{\epsilon})({\bf
p}^{2}{\bf w}^{2}-({\bf p}{\bf w})^{2}),
\end{equation}where the normalization $w^{2}=w_{0}^{2}-{\bf w}^{2}=1$
has been used. It follows from the Schwarz inequality that (53) is
non-negative if the condition (52) is satisfied. If $a^{2}\leq 0$
then we choose the Lorentz frame such that $w=(1,{\bf 0})$. In this
frame
\begin{equation} a_{\mu}a_{\nu}\alpha^{\mu\nu}=
(\epsilon-\pi_{\epsilon})p^{-2}(a_{0}^{2}{\bf p}^{2}+p_{0}^{2}{\bf
a}^{2}-2{\bf p}{\bf a}p_{0}a_{0}) +2\pi_{\epsilon}p^{-2}({\bf
p}^{2}{\bf a}^{2}-({\bf p}{\bf a})^{2}).
\end{equation}
(54) is non-negative under the condition (52) again by Schwarz
inequality. It follows from eqs.(53)-(54) that the conditions (52)
are sufficient and necessary for a definition of the diffusions
(48) and (50) ( the condition (51) is necessary for any definition
of a diffusion because the diffusion matrix $\alpha^{\mu\nu}$
should be equal to the correlation function of particle's
diffusing paths which is defined by a positive probability
measure).

We find the equilibrium solution $\Omega_{E}$ from the requirement
\begin{equation}
(\alpha^{\mu\nu}\partial_{\nu}+r P^{\mu\nu}w_{\nu})\Omega_{E}=0.
 \end{equation} There is a solution (which is  the
J\"uttner equilibrium \cite{jut}), if
\begin{equation}
 r=\beta (\epsilon-\pi_{\epsilon}).
 \end{equation}Then,
\begin{equation}
\Omega_{E}=\exp(-\beta wp-\gamma p^{2}),
\end{equation}
where $\gamma$ is an arbitrary constant. Eq.(56) can be satisfied
for $r=0$ if $\pi_{\epsilon}=\epsilon $ which is still in agreement
with eq.(52). However, the condition $\pi_{\epsilon}=\epsilon $ is
in conflict with the positivity of the measure determining the
correlation functions of the electromagnetic field ( see eq.(32)).

 If the equilibrium $\Omega_{E}$ is known
then the form of the friction can be determined  from the condition
of the detailed balance (see an interpretation in terms of
transition rates in \cite{risken} and as Kolmogorov's reversibility
condition in \cite{kolmogorov})
\begin{equation} \int dp\Omega_{E}\Phi_{\tau}W =\int
dp\Omega_{E}\Phi W_{\tau}.
\end{equation}
 This
condition is equivalent to the requirement that the generator of
the diffusion is of the form
\begin{equation}
{\cal
G}_{w}=\frac{1}{\tau_{c}}p^{\mu}\partial_{\mu}^{x}+\Omega_{E}^{-1}\partial_{\mu}\alpha^{\mu\nu}\Omega_{E}\partial_{\nu}
\end{equation}
 From eq.(59) it follows that the
friction term of a reversible diffusion must have the form
\begin{equation}
\alpha^{\mu\nu}(\Omega_{E}^{-1}\partial_{\mu}\Omega_{E})\partial_{\nu}.
\end{equation}
Comparing eq.(60) with eq.(48) we can see that the diffusion (48)
satisfies the detailed balance condition (58)if $r$ is determined by
eq.(56).
\section{Discussion and summary}
We  have discussed Liouville description of a particle in an
electromagnetic field coming from a thermal radiation. In such a
system there is a distinguished reference frame associated with
the heat bath. The Poincare invariance of the whole system
(particle plus the heat bath) can be expressed as a covariance
with respect to the transformations of the reference frame of an
observer. In such a description the thermal field becomes a
function of the space-time point and the reference frame. We have
shown that the lowest order term in the expansion of the Liouville
evolution in powers of the random electromagnetic field is
described by a diffusion operator which is a generalization of the
relativistic diffusion of Schay and Dudley. We developed a
diffusion theory assuming (after Kubo) that the generator of the
short time dynamics after multiplication by the environmental
correlation time determines the diffusion at long time.  We treat
the square of the momentum as a continuous variable. The off shell
formulation allows to preserve explicit Lorentz covariance. It may
have applications to quantum field theory (heavy ion collisions)
and cosmology (diffusion of galactic matter). It seems that the
diffusion equation derived from an interaction with a random
electromagnetic field could also be inferred on the basis of
general assumptions of the preservation of the square of the
momentum and the Poincare invariance of a system described by the
phase point $ (x,p)$ and a fixed vector $w$ which can be
interpreted as the velocity of a frame in which the particle is
studied.Demanding that similarly as in the case of the
deterministic dynamics the probability distribution does not
depend on the proper time we obtain a diffusive transport equation
in the laboratory time. A similar transport equation (called a
relativistic Ornstein-Uhlenbeck process) has been discussed in
refs.\cite{malik}-\cite{chev2} without relating it to the physical
model of an interaction with an electromagnetic field. The authors
in \cite{malik}-\cite{chev2} investigate diffusion  of particles
in a fluid which supplies a preferred reference frame. However,
their diffusion does not coincide with the one defined in this
paper. Owing to the dependence of $\alpha^{\mu\nu}$ in eq.(50) on
the momenta there is no choice of parameters and no choice of the
frame which enables to reduce the diffusion (50) to the
Ornstein-Uhlenbeck diffusion (the Ornstein-Uhlenbeck diffusion
appears only in the non-relativistic limit).
 It is reassuring that  a mathematical
construction of the diffusion, discovered first by Schay
\cite{schay} and Dudley \cite{dudley} and  generalized here to
realistic systems at finite temperature, is the one which may be
realized in nature (in particular, in a particle's interaction with
a thermal radiation).

\end{document}